\documentstyle[12pt]{article}
\def \beq{\begin{equation}}
\def \eeq{\end{equation}}
\def \ite{{\it et al.}}
\begin{document}
\rightline{CERN-TH/97-227}
\rightline{EFI 97-39}
\rightline{September 1997}
\rightline{hep-ph/9709223}
\bigskip

\centerline{\bf $B$ DECAYS TO CHARMLESS $VP$ FINAL STATES\footnote{To be
submitted to Phys.~Rev.~D.}}
\bigskip
\centerline{\it Amol S. Dighe}
\centerline{\it Enrico Fermi Institute and Department of Physics,
University of Chicago}
\centerline{\it 5640 S. Ellis Avenue, Chicago IL 60615}
\smallskip
\centerline{and}
\smallskip
\centerline{\it Michael Gronau\footnote{Permanent Address: Physics
Department,
Technion -- Israel Institute of Technology, 32000 Haifa, Israel}}
\centerline{\it Theoretical Physics Division, CERN}
\centerline{\it CH-1211 Geneva 23, Switzerland}
\smallskip
\centerline{and}
\smallskip
\centerline{\it Jonathan L. Rosner}
\centerline{\it Enrico Fermi Institute and Department of Physics,
University of Chicago}
\centerline{\it 5640 S. Ellis Avenue, Chicago IL 60615}
\bigskip

\centerline{\bf ABSTRACT}
\bigskip

\begin{quote}
The CLEO Collaboration has now observed the decays $B^+ \to \omega \pi^+$ and
$B \to \omega K^+$ with branching ratios of $(1.1^{+0.6}_{-0.5} \pm 0.2) \times
10^{-5}$ and $(1.5^{+0.7}_{-0.6} \pm 0.3) \times 10^{-5}$, respectively.  These
are the first reported decays to charmless final states involving a vector (V)
and a pseudoscalar (P) meson. The implications of these decays for others of
$B$ mesons to charmless VP final states are explored.  In a model-independent
approach, using only flavor SU(3) symmetry, several tests are proposed for an
anticipated hierarchy among different contributions to decay amplitudes.
\end{quote}
\bigskip
\leftline{PACS numbers: 13.25.Hw, 14.40.Nd, 11.30.Er, 12.15.Ji}

\vfill
\leftline{CERN-TH/97-227}
\leftline{September 1997}
\newpage

\centerline{\bf I.  INTRODUCTION}
\bigskip

The CLEO Collaboration \cite{CLEOVP} has now observed the decays $B^+ \to
\omega \pi^+$ and $B \to \omega K^+$ with branching ratios of
$(1.1^{+0.6}_{-0.5} \pm 0.2) \times 10^{-5}$ (2.9$\sigma$) and
$(1.5^{+0.7}_{-0.6} \pm 0.3) \times 10^{-5}$ (4.3$\sigma$), respectively. 
These are the first reported decays of $B$ mesons to charmless final states
involving a vector ($V$) and a pseudoscalar ($P$) meson.  $VP$ final states may
be crucial in studies of CP violation in $B$ decays \cite{NQ}. 

We have previously applied flavor SU(3) symmetry \cite{Zepp,SW,Chau} to decays
of the form $B \to PP$ \cite{GHLR,GHLRS,GHLRP,Quad,DGRPL,ASD,etap,Vtd}, and
made some preliminary remarks about $VP$ decays in Refs.~\cite{GHLRP} and
\cite{Vtd}. In the latter paper relations are defined between SU(3) amplitudes
and quark diagrams for $VP$ decays. The observation of the $\omega \pi^+$ and
$\omega K^+$ modes, and the existence of limits on other $VP$ modes at levels
close to those expected \cite{CLEOJS}, make an updated analysis relevant at
this time.  The $2.9 \sigma$ level of the $\omega \pi^+$ signal requires that
we regard it as preliminary.

We decompose amplitudes for $B \to VP$ decays into linear combinations of
reduced matrix elements in Section II.  Applications of the relations implied
by these decompositions, suggesting a variety of tests for an anticipated
hierarchy among different contributions, are discussed in Section III. Our
results are compared with attempts to calculate decay modes {\it a priori} with
the help of specific models in Section IV.  We conclude (with a brief summary
of experimental prospects) in Section V. 
\bigskip

\centerline{\bf II.  SU(3) DECOMPOSITION}
\bigskip

In Tables 1 and 2 we list the $VP$ modes of nonstrange $B$ mesons for
strangeness-preserving and strangeness-changing decays, respectively.  Our
notation is as follows: 

\begin{enumerate}

\item As a language equivalent to flavor SU(3), we employ an overcomplete set
of quark diagrams \cite{Chau}, which we denote by $T$ (tree), $C$
(color-suppressed), $P$ (QCD-penguin), $S$ (additional penguin contribution
involving flavor-SU(3)-singlet mesons, called $P_1$ in Ref.~\cite{DGRPL}), $E$
(exchange), $A$ (annihilation) and $PA$ (penguin annihilation). The last three
amplitudes, in which the spectator quark enters into the decay Hamiltonian, are
expected to be suppressed by $f_B/m_B$ ($f_B\approx 180~{\rm MeV}$) and may be
neglected to a good approximation. The presence of higher-order electroweak
penguin contributions \cite{EWP} introduces no new SU(3) amplitudes, and in 
terms of quark graphs merely leads to a substitution \cite{GHLRP,DGRPL} 
$$
T \to t \equiv T + P^C_{EW}~~,~~~
C \to c \equiv C + P_{EW}~~,
$$
\beq \label{eqn:combs}
P\to p \equiv P - {1 \over 3}P^C_{EW}~~,~~~
S \to s \equiv S - {1 \over 3}P_{EW}~~,
\eeq
where $P_{EW}$ and $P^C_{EW}$ are color-favored and color-suppressed
electroweak penguin amplitudes.

\item We use the phase conventions of Ref.~\cite{GHLR} for pseudoscalar mesons,
the mixing assumption $\eta = (s \bar s - u \bar u - d \bar d)/\sqrt{3}$ and
$\eta' = (u \bar u + d \bar d + 2 s \bar s)/\sqrt{6}$, and the corresponding
phase conventions for vector mesons with $\omega = (u \bar u + d \bar d)/
\sqrt{2}$ and $\phi = s \bar s$.

\item We denote strangeness-preserving ($\Delta S = 0$) amplitudes by unprimed
letters and strangeness-changing ($|\Delta S| = 1$) amplitudes by primed
letters. 

\item The suffix on each amplitude denotes whether the spectator quark is
included in a pseudoscalar ($P$) or vector ($V$) meson.

\item Each decay amplitude involves positive or negative integer coefficients
multiplying the indicated reduced amplitudes and divided by a common
denominator factor.

\end{enumerate}

\begin{table}
\caption{$\Delta S = 0$ $B \to VP$ decays. Coefficients of amplitudes are to be
divided by denominator factor.} 
\begin{center}
\begin{tabular}{|r|c|c c|c c|c c|c c|} \hline
& Denom. & $t_P$ & $t_V$ & $c_P$ & $c_V$ & $p_P$ & $p_V$ & $s_P$ & $s_V$ \\
\hline
$B^+ \to \rho^+ \pi^0$ & $-\sqrt{2}$ & 1 & & & 1 & 1 & $-1$ & & \\
$\rho^0 \pi^+$ & $-\sqrt{2}$ & & 1 & 1 & & $-1$ & 1 & & \\
$\omega \pi^+$ & $\sqrt{2}$ & & 1 & 1 & & 1 & 1 & 2 & \\
$\phi \pi^+$ & 1 & & & & & & & 1 & \\
$\rho^+ \eta$ & $-\sqrt{3}$ & 1 & & & 1 & 1 & 1 & & 1 \\
$\rho^+ \eta'$ & $\sqrt{6}$ & 1 & & & 1 & 1 & 1 & & 4 \\
$K^{*+} \bar K^0$ & 1 & & & & & & 1 & & \\
$\bar K^{*0} K^+$ & 1 & & & & & 1 & & & \\ \hline
$B^0 \to \rho^- \pi^+$ & $-1$ & & 1 & & & & 1 & & \\
$\rho^+ \pi^-$ & $-1$ & 1 & & & & 1 & & & \\
$\rho^0 \pi^0$ & 2 & & & $-1$ & $-1$ & 1 & 1 & & \\
$\omega \pi^0$ & 2 & & & 1 & $-1$ & 1 & 1 & 2 & \\
$\phi \pi^0$ & $\sqrt{2}$ & & & & & & &  1 & \\
$\rho^0 \eta$ & $-\sqrt{6}$ & & & $-1$ & 1 & 1 & 1 & & 1 \\
$\rho^0 \eta'$ & $2 \sqrt{3}$ & & & $-1$ & 1 & 1 & 1 & & 4 \\
$\omega \eta$ & $- \sqrt{6}$ & & & 1 & 1 & 1 & 1 & 2 & 1 \\
$\omega \eta'$ & $2 \sqrt{3}$ & & & 1 & 1 & 1 & 1 & 2 & 4 \\
$\phi \eta$ & $-\sqrt{3}$ & & & & & & & 1 & \\
$\phi \eta'$ & $\sqrt{6}$ & & & & & & & 1 & \\
$K^{*0} \bar K^0$ & 1 & & & & & & 1 & & \\
$\bar K^{*0} K^0$ & 1 & & & & & 1 & & & \\ \hline
\end{tabular}
\end{center}
\end{table}

\begin{table}
\caption{$|\Delta S| = 1$ $B \to VP$ decays. Coefficients of amplitudes are to
be divided by denominator factor.} 
\begin{center}
\begin{tabular}{|r|c|c c|c c|c c|c c|} \hline
& Denom. & $t'_P$ & $t'_V$ & $c'_P$ & $c'_V$ & $p'_P$ & $p'_V$ & $s'_P$
& $s'_V$ \\
\hline
$B^+ \to \rho^+ K^0$ & 1 & & & & & & 1 & & \\
$\rho^0 K^+$ & $-\sqrt{2}$ & & 1 & 1 & & & 1 & & \\
$K^{*0} \pi^+$ & 1 & & & & & 1 & & & \\
$K^{*+} \pi^0$ & $-\sqrt{2}$ & 1 & & & 1 & 1 & & & \\
$\omega K^+$ & $\sqrt{2}$ & & 1 & 1 & & & 1 & 2 & \\
$\phi K^+$ & 1 & & & & & 1 & & 1 & \\
$K^{*+} \eta$ & $-\sqrt{3}$ & 1 & & & 1 & 1 & $-1$ & & 1 \\
$K^{*+} \eta'$ & $\sqrt{6}$ & 1 & & & 1 & 1 & 2 & & 4 \\ \hline
$B^0 \to \rho^- K^+$ & $-1$ & & 1 & & & & 1 & & \\
$\rho^0 K^0$ & $\sqrt{2}$ & & & $-1$ & & & 1 & & \\
$K^{*+} \pi^-$ & $-1$ & 1 & & & & 1 & & & \\
$K^{*0} \pi^0$ & $\sqrt{2}$ & & & & $-1$ & 1 & & & \\
$\omega K^0$ & $\sqrt{2}$ & & & 1 & & & 1 & 2 & \\
$\phi K^0$ & 1 & & & & & 1 & & 1 & \\
$K^{*0} \eta$ & $-\sqrt{3}$ & & & & 1 & 1 & $-1$ & & 1 \\
$K^{*0} \eta'$ & $\sqrt{6}$ & & & & 1 & 1 & 2 & & 4 \\ \hline
\end{tabular}
\end{center}
\end{table}
\bigskip

\centerline{\bf III.  APPLICATIONS}
\bigskip

\leftline{\bf A.  Hierarchies of amplitudes}
\bigskip

One can immediately identify certain amplitudes likely to be most important in
$B \to VP$ decays. In the corresponding $PP$ decays (which may be denoted by
similar amplitudes without the subscripts), the amplitudes are expected to obey
an approximate hierarchy \cite{GHLR,GHLRS,GHLRP,etap,SilWo}.  The process $B^0
\to K^+ \pi^-$ is observed with a branching ratio somewhat in excess of
$10^{-5}$, while $B^0 \to \pi^+ \pi^-$ is expected to have a branching ratio
not vastly different from this.  Thus we deduced in previous work that $|t|
\simeq |p'|$, while $|p/t| \simeq |t'/p'| \simeq \lambda$, where $\lambda
\equiv V_{us} \simeq 0.22$.  We do not have an estimate for $|c/t|$ or
$|c'/t'|$.  We expect $|c/t|$ to be small on the basis of color-suppression
arguments. However, $|c'/t'|$ may be larger due to the electroweak penguin term
in $c'$ [see Eq.~(1)]. The large branching ratios for $B^+ \to K^+ \eta'$
(about $7 \times 10^{-5}$) and $B^0 \to K^0 \eta'$ (about $5 \times 10^{-5}$)
\cite{CLEOetap} indicate the importance of the $s'$ amplitude at a level
comparable to that of $p'$ \cite{etap,HJL}. 

A similar hierarchy appears to apply to the $VP$ decays.  The fact that the 
$B^+ \to \omega \pi^+$ and $B^+ \to \omega K^+$ branching ratios are comparable
to one another and each of order $10^{-5}$ indicates that the dominant
contribution to $\omega \pi^+$ is most likely $t_V$, while the dominant
contribution to $\omega K^+$ is most likely $p'_V$.  We expect the $s'_P$
contribution to be relatively unimportant; this contribution would involve
a coupling of the $\omega$ and $\phi$ which violated the Okubo-Iizuka-Zweig
(OZI) rule favoring connected quark diagrams.  Such couplings are probably much
more important for $\eta$ and $\eta'$ than for vector mesons.  Specifically,
the penguin amplitude $s'_V$, coupling to the flavor SU(3) singlet component of
the $\eta$ and and $\eta'$, can be as large as or even larger than $p'_V$. A
similar situation seems to hold in decays to two light pseudoscalar mesons
\cite{etap}. 
% \bigskip
\newpage

\leftline{\bf B.  Tests for smallness of amplitudes}
\bigskip

How can one learn more about which $VP$ amplitudes are important? One way of
using the tables is to compare charged $B$ decays and neutral $B$ decays to
each other.  This can teach us something about the magnitudes of some of the
amplitudes. Consider, for instance, the eight pairs of $|\Delta S| = 1$
processes listed in Table 2.  The following approximate amplitude equalities 
test the smallness of certain contributions. The relations between $B^+$ and 
$B^0$ amplitudes are independent of SU(3) breaking. In each case we list only
the final state. 

\begin{enumerate}

\item Smallness of $c'_{P,V}$:

$$
\rho^+ K^0 \approx \sqrt{2} (\rho^0 K^0)~~,~~~
\sqrt{2}(\rho^0 K^+) \approx \rho^- K^+~~~,
$$
\beq
K^{*0} \pi^+ \approx \sqrt{2} (K^{*0} \pi^0)~~,~~~
\sqrt{2}(K^{*+} \pi^0) \approx K^{*+} \pi^-~~~.
\eeq
The $c'_{P,V}$ amplitudes contain color-favored electroweak penguin terms which
may not be negligible (see Eq.~(1) and Ref.~\cite{EWP}), and indeed provide
important contributions in {\it a priori} calculations to be discussed in
Sec.~IV. 

\item Smallness of $t'_{P,V}$:

\beq
\omega K^+ \approx \omega K^0~~,~~~
K^{*+} \eta \approx K^{*0} \eta~~,~~~
K^{*+} \eta' \approx K^{*0} \eta'~~~.
\eeq

\item Smallness of $s'_P$:

\beq
-\rho^0 K^+ \approx \omega K^+~~,~~~
K^{*0} \pi^+ \approx \phi K^+~~~.
\eeq
The first relation is sensitive to any breakdown of nonet symmetry (unequal
decay constants for $\rho$ and $\omega$ mesons).  The second relation is
sensitive to SU(3)-breaking effects since it involves comparing an amplitude
with nonstrange quark pair production to one with strange quark pair
production; the form factors are also likely to differ \cite{Vtd}. 

In addition, a number of approximate triangle relations hold, such as
\beq
\sqrt{2} (\rho^+ K^0) \approx \rho^0 K^0 + \omega K^0~~~,
\eeq
all of whose sides have a $p'_V$ contribution, so the decay rates may be
significant.  The shape of the amplitude triangle may tell us about the
relative magnitudes and phases of $c'_P$ and $p'_V$.  Since we expect $c'_P$ to
be smaller than $p'_V$, this triangle will be a ``squashed'' one.

\item The last relation among the eight pairs of amplitudes, $\phi K^+ =
\phi K^0$, is exact and follows from isospin.

\end{enumerate}

Assuming that both $s'_P$ and $c'_P$ are small, one also finds $\rho^0 K^0
\approx \omega K^0$. Lipkin \cite{HJL} has pointed out that if $s'_P$ is small
[as checked, for example, by relations such as Eq.~(4)], but if the $B^0 \to
\rho^0 K^0$ and $B^0 \to \omega K^0$ rates are {\it unequal}, then both
$p'_V$ and $c'_P$ amplitudes must be present, and they are close enough in
amplitude for interference to be observed.  Whether this enhances the
possibility of observing direct CP violation \cite{HJL} remains an open
question. The most likely source of a contribution to $c'_P$ is an electroweak
penguin amplitude with the same weak phase as $p'_V$, so a direct CP asymmetry
is unlikely in the neutral decays.  In the more readily observed charged decays
$B^+ \to (\rho^0, \omega) K^+$, one would need interference between $t'_P$ and
$p'_V$ to see a CP asymmetry.  In our approach one cannot infer anything about
$t'_V$ from $c'_P$. 

The $\Delta S = 0$ amplitudes do not exhibit simple isospin relations
which test the smallness of some amplitudes.  Still, one can make the
following observations about large and small amplitudes:

\begin{enumerate}

\item The largest amplitudes are expected to be $t_P,~t_V$.  Therefore,
the following 7 processes are expected to have the largest rates:
$\rho^+ \pi^0$, $\rho^0 \pi^+$, $\omega \pi^+$, $\rho^+ \eta$,
$\rho^+\eta'$, $\rho^- \pi^+$, and $\rho^+ \pi^-$.

\item Smaller decay rates (equal in $B^+$ and $B^0$ decays) measure
different kinds of penguin amplitudes:

\begin{itemize}

\item $p_P$ is measured in $\bar K^{*0} K^+ = \bar K^{*0} K^0$;

\item $p_V$ is measured in $K^{*+} \bar K^0 = K^{*0} \bar K^0$;

\item $s_P$ is measured in $\phi \pi^+ = \sqrt{2}(\phi \pi^0) = - \sqrt{3}
(\phi \eta) = \sqrt{6} (\phi \eta')$;

\item $s_V$ is measured by the combinations $\sqrt{6}(\rho^+ \eta') +
\sqrt{3}(\rho^+ \eta)$, $2 \sqrt{3}(\rho^0 \eta') + \sqrt{6}(\rho^0 \eta)$,
and $2 \sqrt{3}(\omega \eta') + \sqrt{6}(\omega \eta)$, implying rate
relations if $s_V$ is small.

\end{itemize}

\end{enumerate} 
\bigskip

\leftline{\bf C.  Relations testing for presence or absence of $s'_V$}
\bigskip

Whereas the amplitude $p'_V$ can be measured directly in $B^+ \to \rho^+ K^0$,
it is much more difficult to determine the amplitude $s'_V$ contributing only
to decays involving the $\eta$ and $\eta'$.  We recall that the presence of the
corresponding sizable amplitude $s'$ in $B \to PP$ decays was manifested by the
particularly large $B \to K \eta'$ decay rate \cite{etap}. 

Several tests for the presence of the singlet amplitude $s'_V$ can be
constructed. $s'_V$ is less likely to be small than $s'_P$, since the axial
anomaly can affect $\eta$ and $\eta'$ flavor-singlet couplings \cite{anomaly}.
A large corresponding singlet amplitude in $|\Delta S| = 1$ $B \to PP$ decays
was found to be needed to explain the observed $B \to K \eta'$ rate
\cite{etap}. 

In the event that $s'_V$ {\em is} negligible, several triangle relations hold
among the amplitudes for $B^+ \to \rho^0 K^+$ and $B^+ \to K^{*+} (\pi^0, \eta,
\eta')$, and among the amplitudes for $B^0 \to K^{*0} (\eta,\eta')$ and $B^0
\to \rho^+ K^0$. If we are willing, moreover, to neglect $c'_V$ in comparison
with other amplitudes, a set of triangle relations analogous to the above, but
with the decays $B^0 \to K^{*0} (\eta,\eta')$ replacing $B^+ \to K^{*+}
(\eta,\eta')$ and with $K^{*0} \pi^+$ replacing $-\sqrt{2}(K^{*+} \pi^0)$,
should hold.

Under some circumstances, interference terms between contributions of different
amplitudes to rates will cancel when suitable sums of rates are constructed.
Thus, one finds that if $s'_V$ can be neglected, the triangle relations
mentioned above imply 
\beq
{\cal B}(B^+ \to K^{*+} \pi^0) + {\cal B}(B^+ \to \rho^+ K^0)
= {\cal B}(B^+ \to K^{*+} \eta) + {\cal B}(B^+ \to K^{*+} \eta')~~~.
\eeq
Similarly, with $c'_V$ small in addition, one finds
\beq
{\cal B}(B^+ \to K^{*0} \pi^+)/2 + {\cal B}(B^+ \to \rho^+ K^0)
= {\cal B}(B^0 \to K^{*0} \eta) + {\cal B}(B^0 \to K^{*0} \eta')~~~.
\eeq
(Here and later we neglect phase space effects.) Both sides of these two
relations contain contributions from the $p'_V$ amplitude, which we expect to
be significant.  The failure of either of the above two equations to hold would
indicate a significant $s'_V$ contribution. In that case, we may proceed to
determine $s'_V$ as follows. 

When the amplitude $s'_V$ is not neglected, the triangle relations for
$B^+$ decays discussed above are replaced by
\beq
-\sqrt{2}(K^{*+} \pi^0) - (\rho^+ K^0) + s'_V = -\sqrt{3}(K^{*+} \eta)~~~,
\eeq
\beq
-\sqrt{2}(K^{*+} \pi^0) + 2 (\rho^+ K^0) + 4 s'_V = \sqrt{6}(K^{*+} \eta')~~~.
\eeq
Moreover, if $c'_V$ can be neglected, one can write
\beq
(K^{*0} \pi^+) - (\rho^+ K^0) + s'_V = -\sqrt{3}(K^{*0} \eta)~~~,
\eeq
\beq
(K^{*0} \pi^+) + 2 (\rho^+ K^0) + 4s'_V = \sqrt{6}(K^{*0} \eta')~~~.
\eeq
In the complex plane, let $(\rho^+ K^0) = (1,0)$ [so that all the amplitudes
and phases in these relations are measured in units of $(\rho^+ K^0)$].  
Since only $p'_V$ contributes to $(\rho^+ K^0)$, all the amplitudes will
henceforth be given in terms of $p'_V$ as the unit.

Let $-\sqrt{2}(K^{*+} \pi^0) = (a,b)$, $(K^{*0} \pi^+) = (c,d)$, and $s'_V =
(e,f)$.  Then, squaring the above four equations to obtain four rate relations,
and constraining $a^2 + b^2$ and $c^2 + d^2$ using the rates for the decays
$B^+ \to K^{*+} \pi^0$ and $B^+ \to K^{*0} \pi^+$, we have six equations in the
six unknowns $a,b,c,d,e,f$.  One is required to measure seven different decay
rates, but all of them involve $p'_V$, so none of them should be very small. 
\bigskip

\leftline{\bf D.  Where does the spectator quark end up?}
\bigskip

The distinction between amplitudes $a_P$ or $a'_P$ ($a \equiv t,~c,~p,~s$) (in
which the spectator quark is incorporated into a pseudoscalar meson) and $a_V$
or $a'_V$ (in which the spectator ends up in a vector meson) is responsible for
the large number of reduced amplitudes in the $VP$ case, as compared to the
simpler $PP$ decays.  Some hint that these amplitudes may not have equal
magnitudes is provided by the upper bound \cite{CLEOVP} ${\cal B}(B^+ \to \phi
K^+) < 0.53 \times 10^{-5}$, as compared with ${\cal B}(B^+ \to \omega K^+) =
(1.5^{+0.7}_{-0.6} \pm 0.3) \times 10^{-5}$, implying ${\cal B}(B^+ \to \phi
K^+)/{\cal B}(B^+ \to \omega K^+) < 1$.  The $\phi K^+$ amplitude is dominantly
$p'_P$, while the $\omega K^+$ amplitude is mainly $p'_V/\sqrt{2}$.  If the
$p'_P$ and $p'_V$ amplitudes were equal, we should have expected ${\cal B}(B^+
\to \phi K^+) = 2 {\cal B}(B^+ \to \omega K^+)$. The amplitude $p'_V$ can be
measured directly in $B^+ \to \rho^+ K^0$ and, neglecting $c'_P$, in $B^0 \to
\rho^0 K^0$.  To confirm that indeed $|p'_V| > |p'_P|$, it would be useful to
compare $p'_V$ measured in this way with $p'_P$ measured in $B^+ \to K^{*0}
\pi^+$. 

The assumption of equal and opposite $p'_P$ and $p'_V$ amplitudes lies behind a
prediction by Lipkin \cite{HJL} that one expects constructive interference
between the nonstrange and strange components of the $\eta$, and destructive
interference in the $\eta'$, for the decays $B^+ \to K^{*+} (\eta,\eta')$. This
is valid if the penguin transition $\bar b \to \bar s$ leads to an intermediate
$\bar s u$ final state accompanied by any number of gluons, as long as there is
not some fundamental asymmetry in the wave function between the $\bar s$ and
the $u$. If the final light $q \bar q$ pair is then produced in a flavor-SU(3)
invariant manner, the $p'_P$ and $p'_V$ amplitudes will be equal and opposite. 
(Gluons must be present; otherwise one could rotate away any $\bar b \to \bar
s$ transition by a redefinition of quark fields \cite{CG}.) 

The full generality of Lipkin's argument is less obvious.  The transition
$\bar b \to \bar s +$ (meson) has different Lorentz structure when the meson
is a pseudoscalar (our $p'_V$ amplitude) than when it is a vector (our $p'_P$
amplitude).  The $p'_V$ and $p'_P$ amplitudes indeed fail to be equal and
opposite in several explicit calculations to be discussed in Sec.~IV.

\begin{table}
\caption{Some $B$ decay modes capable of distinguishing between
$a^{(')}_P$-type and $a^{(')}_V$-type amplitudes.} 
\begin{center}
\begin{tabular}{r c c c c} \hline
Decay & Dominant  & Signal & Expected   & ${\cal B}$ upper \\
mode  & amplitude & events & background & limit $\times 10^{-6}$ \\ \hline
$B^+ \to \rho^+ \pi^0$ & $t_P$  & 8 & $5.5 \pm 1.2$ & 77 \\
        $\rho^0 \pi^+$ & $t_V$  & 4 & $2.3 \pm 0.3$ & 43 \\
        $\omega \pi^+$ & $t_V$  & 8.8 & & (a) \\
        $\phi \pi^+$   & $s_P$  & 0 & & 5.6 \\
        $\rho^+ K^0$   & $p'_V$ & 0 &      0        & 48 \\
        $\rho^0 K^+$   & $p'_V$ & 1 & $3.8 \pm 0.2$ & 19 \\
        $\omega K^+$   & $p'_V$ & 12 & & (a) \\
        $\phi K^+$     & $p'_P$ & 0 & & 5.3 \\
        $K^{*0} \pi^+$ & $p'_P$ & 2 & $1.0 \pm 0.6$ & 41 \\
        $K^{*+} \pi^0$ & $p'_P$ & 4 & $1.9 \pm 0.7$ & 99 \\ \hline
$B^0 \to \rho^- \pi^+$ & $t_V$  & (b) &    (b)     & (c) \\
        $\rho^+ \pi^-$ & $t_P$  & (b) &    (b)     & (c) \\
        $\phi \pi^0$   & $s_P$  & 0 & & 6.5 \\
        $\rho^- K^+$   & $p'_V$ & 2 & $2.0 \pm 0.4$ & 35 \\
        $\rho^0 K^0$   & $p'_V$ & 0 &      0        & 39 \\
        $\phi K^0$     & $p'_P$ & 2 & & 42 \\
        $K^{*+} \pi^-$ & $p'_P$ & 3 & $0.7 \pm 0.2$ & 72 \\
        $K^{*0} \pi^0$ & $p'_P$ & 0 & $1.1 \pm 0.3$ & 28 \\ \hline
\end{tabular}
\end{center}
\leftline{(a) Signal observed (see text).}
\leftline{(b) Sum of channels has 7 events above expected background of
$2.9 \pm 0.7$.}
\leftline{(c) Upper limit on average branching ratio is $88 \times 10^{-6}$.}
\end{table}

One might ask whether there is {\it any} evidence so far for amplitudes of the
$a^{(')}_P$ type.  In Table 3 we collect a number of decay modes which can shed
light on this question.  We list the mode, the amplitude expected to dominate,
and the number of signal events, expected backgrounds, and upper limits on the
branching ratios in units of $10^{-6}$ reported by CLEO II \cite{CLEOJS}.  We
do not list the coefficients of the dominant amplitudes, which may be found in
Tables 1 and 2. 

None of the upper limits shown in Table 3 conflicts with the expectation that
the $t_V$ amplitude has already been seen in $B^+ \to \omega \pi^+$ and the
$p'_V$ amplitude has already been seen in $B^+ \to \omega K^+$.  If these are
the dominant amplitudes one expects
\beq
{\cal B}(B^0 \to \rho^- \pi^+) = 2 {\cal B}(B^+ \to \rho^0 \pi^+)
= 2 {\cal B}(B^+ \to \omega \pi^+)
\eeq
and
$$
{\cal B}(B^+ \to \rho^+ K^0) = {\cal B}(B^0 \to \rho^- K^+)
$$
\beq
= 2 {\cal B}(B^+ \to \rho^0 K^+) = 2 {\cal B}(B^0 \to \rho^0 K^0)
= 2 {\cal B}(B^+ \to \omega K^+)~~~.
\eeq

The case of $B^0 \to \rho^\pm \pi^\mp$ is particularly interesting since the
excess of signal over expected background is the largest of any in Table 3, but
the process requires flavor tagging in order to separate the $t_V$-dominated
decay $B^0 \to \rho^- \pi^+$ from the $t_P$-dominated decay $B^0 \to \rho^+
\pi^-$.  Interesting statements about the relative magnitudes of amplitudes
will require about a factor of 3 more data than those on which Table 3 was
based. 
\bigskip

\leftline{\bf E.  Relations between $\Delta S = 0$ and $|\Delta S| = 1$ decays}
\bigskip

Comparison of decay rates between $\Delta S = 0$ and $|\Delta S| = 1$ processes
can help in determining ratios of magnitudes of CKM elements.
For example, one expects 
\beq
\frac{{\cal B}(B^+ \to K^{*0} \pi^+)}{{\cal B}(B^+ \to \bar K^{*0} K^+)} =
\frac{{\cal B}(B^+ \to \rho^+ K^0)}{{\cal B}(B^+ \to K^{*+} \bar K^0)}
= |V_{ts}/V_{td}|^2~~~,
\eeq
assuming the top quark dominates the penguin amplitudes in these sets of
processes. In certain ratios SU(3) breaking form factor effects cancel out.
(See Ref.~\cite{Vtd} for a more complete discussion).

Once the dominant amplitudes (such as $t_V$ and $p'_V$, for which we already
have evidence) have been mapped out, one can use flavor SU(3) to anticipate
the smaller amplitudes (like $t'_V$ and $p_V$).  Taking account of SU(3)
breaking, we have
\beq
t'_V/t_V = t'_P/t_P = \lambda(f_K/f_\pi)~~~.
\eeq
For penguin amplitudes in the flavor-SU(3) limit,
\beq
p_V/p'_V = p_P/p'_P = s_V/s'_V = s_P/s'_P =V_{td}/V_{ts}
\eeq
(assuming top-quark dominance of the penguin amplitudes).  One can then, in the
manner of Ref.~\cite{etap}, search for processes in which two amplitudes with
two different weak phases contribute with comparable strength.  If these
amplitudes have different strong phases as well, there is a possibility of a
large CP asymmetry in comparing the rate for a process with its charge
conjugate. 
\bigskip

\centerline{\bf IV.  PREDICTIONS OF SPECIFIC MODELS}
\bigskip

Calculations of $B \to (PP,VP,VV)$ decay rates over the past ten years,
involving assumptions about factorization and using specific $B$-to-light-meson
form factors, include those of Refs.~\cite{Chau,BSW,DT,DeA,KP,Al,CT,Ciu}.  For
the most part, these works have successfully anticipated those $B \to PP$
decays observed at a branching ratio level of $10^{-5}$ or greater, such as
$B^0 \to K^+ \pi^-$. The decay $B^0 \to \pi^+ \pi^-$ is expected to correspond
to a branching ratio not much below $10^{-5}$. 

A common feature of all the calculations is their expectation that ${\cal
B}(B^+ \to \omega K^+)/{\cal B}(B^+ \to \phi K^+) \ll 1$, in disagreement with
the CLEO results \cite{CLEOVP}.  In our language, these calculations predict
$|p'_V/p'_P| \ll 1$; in the work of Ref.~\cite{DT}, the contribution of $p'_V$
is vanishingly small in certain decays. This is a result of the specifically
chosen form factors used to calculate hadronic matrix elements of penguin
operators. 

The authors of Ref.~\cite{Ciu} propose a new source of penguin terms,
associated with charmed quarks in the loop rather than top quarks as is
conventionally assumed (see also Ref.~\cite{BF}).  Their prediction of the
$B^+\to \omega K^+/\phi K^+$ ratio of rates nonetheless remains too small, with
${\cal B} (B^+ \to \omega K^+)/{\cal B}(B^+ \to \phi K^+) \approx 1/6$ in
contradiction to experiment.  This assumption implies weak phases of Arg
$(V^*_{cb} V_{cd}) = \pi$ and Arg $(V^*_{cb} V_{cs}) = 0$ for $\Delta S = 0$
and $|\Delta S| = 1$ penguin amplitudes, respectively. If the top quark were
dominant instead, one would have weak phases of Arg $(V^*_{tb} V_{td}) = -
\beta$ and Arg $(V^*_{tb} V_{ts}) = \pi$.  One cannot distinguish between a
weak phase of $\pi$ and one of zero in the $|\Delta S| = 1$ transitions, but it
should be possible to tell the difference between the weak phase of zero and
$-\beta$ for the $\Delta S = 0$ amplitudes.  That distinction lies beyond the
scope of the present work. 

The relative signs of contributions from $p'_P$ and $p'_V$ in the model of
Ref.~\cite{Chau} are such that one expects ${\cal B}(B \to K^* \eta) > {\cal
B}(B \to K^* \eta')$ (see also Ref.~\cite{HJL}).  For the corresponding decays
with $K$ replacing $K^*$, the prediction \cite{Chau,HJL} is ${\cal B}(B \to K
\eta) < {\cal B}(B \to K \eta')$, as observed for charged $B$'s
\cite{CLEOetap}. 

The authors of Ref.~\cite{Chau} do not include the $s'_V$ terms, which may be
important.  For the corresponding $PP$ decays, numerous authors
\cite{etap,HJL,anomaly} have noted that the singlet amplitude $s'$ is required
to understand the large $B \to K \eta'$ rate.  There is some question as to
the origin of this singlet amplitude.  It most likely originates as a result of
the gluonic anomaly in the axial U(1) current, but may be manifested in various
ways, e.g., through an admixture of $c \bar c$ pairs or gluon pairs in the
$\eta'$ (and, to a lesser extent, $\eta$) wave function.  The absence of a
meaningful constraint on the strange-quark content of the $\eta'$ (achievable
by measuring the rate for $\phi \to \eta' \gamma$) \cite{JLRetap} leaves some
room for such admixtures.  A direct coupling of singlet pseudoscalar mesons
through penguin-type diagrams introduced in Ref.~\cite{Quad} is another
possibility. 

The models do seem to predict roughly the right magnitude for $t_V$, which we
expect to dominate the observed decay $B^+ \to \omega \pi^+$.  An expectation
of Ref.~\cite{Chau} common to other models is that $|t_P| > |t_V|$.
In that case, one should expect ${\cal B}(B^+ \to \rho^+ \pi^0) > {\cal B}(B^+
\to \rho^0 \pi^+)$ and ${\cal B}(B^0 \to \rho^+ \pi^-) > {\cal B}(B^0
\to \rho^- \pi^+)$.  As one sees in Table 3, there is no evidence yet for or
against this hierarchy.

We find that many of our Eqs.~(2)--(4) are not satisfied by the models of Chau
\ite~(Ref.~\cite{Chau}) and Ciuchini \ite~(Ref.~\cite{Ciu}).  The sources of
the violations are of some interest.

In Ref.~\cite{Chau}, the annihilation amplitude seems to have a large effect.
We have noted that the $\phi K^+$ and $\phi K^0$ amplitudes should be equal by
isospin when we neglect annihilation. Chau \ite~find (including annihilation)
that the branching ratios of these processes are $14 \times 10^{-6}$ and $9
\times 10^{-6}$, respectively. This means that the ratio of the annihilation
amplitude to $p'_P$ must be at least 0.25 (which is surprisingly large) {\it
if} $p'_P$ and annihilation interfere constructively in $\phi K^+$. Otherwise,
annihilation is even larger.  Note that in the calculation of Ref.~\cite{Chau}
(as in the others) $p'_P$ (rather than $p'_V$) is the dominant amplitude in
$|\Delta S| = 1$ decays. Thus, the effects of annihilation in other processes
(dominated by the smaller $p'_V$) are even larger.  For instance, these authors
find ${\cal B}(B^+ \to \rho^+ K^0) = 0.34 \times 10^{-6}$ and $ 2{\cal B}(B^0
\to \rho^0 K^0) = 0.70 \times 10^{-6}$, while these numbers should be
approximately equal by our Eq.~(2) if $c'_P$ is negligible. 

In the calculation by Ciuchini \ite~the last two rates come out to be quite
different as a result of a significant electroweak penguin contribution to the
$c'_P$ amplitude. The authors find ${\cal B}(B^+ \to \rho^+ K^0)= 30 \times
10^{-6}$ and $ 2{\cal B}(B^0 \to \rho^0 K^0) = 10 \times 10^{-6}$. The
one-to-two orders of magnitude enhancement relative to the results of
Ref.~\cite{Chau} comes from the ``charming penguin" terms. This illustrates the
large spread of model-predictions. 

\begin{table}
\caption{Branching ratios for penguin-dominated $B$ decays in two models,
in units of $10^{-6}$.}
\begin{center}
\begin{tabular}{c c c} \hline
Decay & Chau \ite & Ciuchini \ite \\ \hline
$B^+ \to \rho^0 K^+$ & 0.6 & 7 \\
$B^+ \to \omega K^+$ & 1.4 & 5.4 \\ \hline
$B^+ \to K^{*0} \pi^+$ & 8.8 & 52 \\
$B^+ \to \phi K^+$ & 14 & 33 \\ \hline
\end{tabular}
\end{center}
\end{table}

In Table 4 we compare some other results from Ref.~\cite{Chau} (see also
Ref.~\cite{CT}) and Ref.~\cite{Ciu}, where again rates are enhanced by
``charming penguins". The first authors neglected $s'_P$. In our treatment, the
branching ratios for the processes in the first and second rows should be
approximately equal; so should those for the processes in the third and fourth
rows.  The differences in Ciuchini \ite~show the effect of $s'_P$ and involve
some nonet-symmetry and SU(3) breaking effects. The combined effect, resulting
in amplitude differences at a level of 20$\%$, is not unexpected. On the other
hand, the rate difference between the first two processes in Chau \ite,
arising from nonet-symmetry breaking alone, seems quite large for such effects. 

\bigskip

\centerline{\bf V.  CONCLUSIONS}
\bigskip

The decays $B^+ \to \omega \pi^+$ (still requiring confirmation) and $\omega
K^+$ seen at branching ratio levels of about $10^{-5}$ by the CLEO
Collaboration \cite{CLEOVP} can be used, with the help of flavor SU(3), to
anticipate the observability of other charmless $B \to VP$ decays in the near
future.  We have indicated which amplitudes in the flavor-SU(3) decomposition
are likely to be large as a result of present evidence. These consist of a
strangeness-preserving ``tree'' amplitude $t_V$ and a strangeness-changing
penguin amplitude $p'_V$.  In both cases the subscript indicates that the
spectator quark is incorporated into a vector ($V$) meson. 

Other decays depending on the amplitude $t_V$ are $B^+ \to \rho^0 \pi^+$ and 
$B^0 \to \rho^- \pi^+$.  If $t_V$ is the dominant amplitude in these processes,
we expect $\Gamma(B^+ \to \rho^0 \pi^+) = \Gamma(B^+ \to \omega \pi^+)$ and
$\Gamma(B^0 \to \rho^- \pi^+) = 2 \Gamma(B^+ \to \omega \pi^+)$.  Furthermore,
model calculations predicting $|t_P| > |t_V|$ imply that decays expected to be
dominated by $t_P$, such as $B^+ \to \rho^+ \pi^0$ and $B^0 \to \rho^+ \pi^-$,
will also have branching ratios in excess of $10^{-5}$. 

An appreciable value for the amplitude $p'_V$, somewhat of a surprise in
conventional models, implies that $B \to \rho K$ decays should be observable at
branching ratio levels in excess of $10^{-5}$.  The smallness of the ratio
${\cal B}(B^+ \to \phi K^+)/{\cal B}(B^+ \to \omega K^+)$ indicates that
$|p'_P| < |p'_V|$.  The amplitude $p'_P$ should dominate not only $B \to \phi
K$ but also $B \to K^* \pi$ decays.  Evidence for any of these would then tell
us the magnitude of $p'_P$.  The relative phase of $p'_P$ and $p'_V$ is probed
by $B \to K^* (\eta,\eta')$ decays. 

We have argued that singlet amplitudes $s,s'$, corresponding to disconnected
quark diagrams, are more likely to be appreciable when pseudoscalar mesons
$\eta,\eta'$ are disconnected from the rest of the diagram than when vector
mesons $\omega,\phi$ are disconnected.  Thus, we expect $|s'_V| > |s'_P|$
and $|s_V| > |s_P|$.  (Recall that the subscript refers to the meson in which
the {\it spectator} quark is incorporated.)  We have suggested several tests
for non-zero singlet amplitudes, including a number of triangle and rate
relations, and have outlined a program for determining the magnitude and
phase of $s'_V$.

Once the dominant $t$, $p'$, and $s'_V$ amplitudes have been determined,
flavor SU(3) predicts the amplitudes $t'$, $p$, and $s_V$.  One can then (cf.
Ref.~\cite{etap}) determine which processes are likely to exhibit noticeable
interferences between two or more amplitudes, thereby having the potential for
displaying direct CP-violating asymmetries.

The CLEO Collaboration \cite{CLEOVP} has also reported the observation of the
decay $B \to \phi K^*$, with a branching ratio of $(1.3^{+0.7}_{-0.6} \pm 0.2)
\times 10^{-5}$ when charged and neutral modes are combined.  (Isospin
invariance implies equal rates for the two.)  Decays of the form $B \to VV$ are
more complicated than $B \to PP$ or $B \to VP$ decays because of the three
possible partial waves in the final state.  Once these are separated out, for
example using decay angular distributions \cite{Ang}, an analysis similar to
the one presented here becomes possible for $B \to VV$ decays as well. 
\bigskip

\centerline{\bf ACKNOWLEDGMENTS}
\bigskip

We are grateful to M. Ciuchini, H. J. Lipkin, M. Neubert, J. G. Smith, and S.
Stone for helpful discussions. This work was performed in part at the Aspen
Center for Physics, and supported in part by the United States Department of
Energy under Grant No.~DE FG02 90ER40560 and by the United States -- Israel
Binational Science Foundation under Research Grant Agreement 94-00253/2. 
\bigskip

% Journal and other miscellaneous abbreviations for references
\def \ajp#1#2#3{Am. J. Phys. {\bf#1}, #2 (#3)}
\def \apny#1#2#3{Ann. Phys. (N.Y.) {\bf#1}, #2 (#3)}
\def \app#1#2#3{Acta Phys. Polonica {\bf#1}, #2 (#3)}
\def \arnps#1#2#3{Ann. Rev. Nucl. Part. Sci. {\bf#1}, #2 (#3)}
\def \art{and references therein}
\def \cmts#1#2#3{Comments on Nucl. Part. Phys. {\bf#1}, #2 (#3)}
\def \cn{Collaboration}
\def \cp89{{\it CP Violation,} edited by C. Jarlskog (World Scientific,
Singapore, 1989)}
\def \dpfa{{\it The Albuquerque Meeting: DPF 94} (Division of Particles and
Fields Meeting, American Physical Society, Albuquerque, NM, Aug.~2--6, 1994),
ed. by S. Seidel (World Scientific, River Edge, NJ, 1995)}
\def \dpff{{\it The Fermilab Meeting: DPF 92} (Division of Particles and Fields
Meeting, American Physical Society, Batavia, IL., Nov.~11--14, 1992), ed. by
C. H. Albright \ite~(World Scientific, Singapore, 1993)}
\def \efi{Enrico Fermi Institute Report No. EFI}
\def \epl#1#2#3{Europhys.~Lett.~{\bf #1}, #2 (#3)}
\def \f79{{\it Proceedings of the 1979 International Symposium on Lepton and
Photon Interactions at High Energies,} Fermilab, August 23-29, 1979, ed. by
T. B. W. Kirk and H. D. I. Abarbanel (Fermi National Accelerator Laboratory,
Batavia, IL, 1979}
\def \hb87{{\it Proceeding of the 1987 International Symposium on Lepton and
Photon Interactions at High Energies,} Hamburg, 1987, ed. by W. Bartel
and R. R\"uckl (Nucl. Phys. B, Proc. Suppl., vol. 3) (North-Holland,
Amsterdam, 1988)}
\def \ib{{\it ibid.}~}
\def \ibj#1#2#3{~{\bf#1}, #2 (#3)}
\def \ichep72{{\it Proceedings of the XVI International Conference on High
Energy Physics}, Chicago and Batavia, Illinois, Sept. 6 -- 13, 1972,
edited by J. D. Jackson, A. Roberts, and R. Donaldson (Fermilab, Batavia,
IL, 1972)}
\def \ijmpa#1#2#3{Int. J. Mod. Phys. A {\bf#1}, #2 (#3)}
\def \jpb#1#2#3{J.~Phys.~B~{\bf#1}, #2 (#3)}
\def \lkl87{{\it Selected Topics in Electroweak Interactions} (Proceedings of
the Second Lake Louise Institute on New Frontiers in Particle Physics, 15 --
21 February, 1987), edited by J. M. Cameron \ite~(World Scientific, Singapore,
1987)}
\def \ky{{\it Proceedings of the International Symposium on Lepton and
Photon Interactions at High Energy,} Kyoto, Aug.~19-24, 1985, edited by M.
Konuma and K. Takahashi (Kyoto Univ., Kyoto, 1985)}
\def \mpla#1#2#3{Mod. Phys. Lett. A {\bf#1}, #2 (#3)}
\def \nc#1#2#3{Nuovo Cim. {\bf#1}, #2 (#3)}
\def \np#1#2#3{Nucl. Phys. {\bf#1}, #2 (#3)}
\def \pisma#1#2#3#4{Pis'ma Zh. Eksp. Teor. Fiz. {\bf#1}, #2 (#3) [JETP Lett.
{\bf#1}, #4 (#3)]}
\def \pl#1#2#3{Phys. Lett. {\bf#1}, #2 (#3)}
\def \pla#1#2#3{Phys. Lett. A {\bf#1}, #2 (#3)}
\def \plb#1#2#3{Phys. Lett. B {\bf#1}, #2 (#3)}
\def \pr#1#2#3{Phys. Rev. {\bf#1}, #2 (#3)}
\def \prc#1#2#3{Phys. Rev. C {\bf#1}, #2 (#3)}
\def \prd#1#2#3{Phys. Rev. D {\bf#1}, #2 (#3)}
\def \prl#1#2#3{Phys. Rev. Lett. {\bf#1}, #2 (#3)}
\def \prp#1#2#3{Phys. Rep. {\bf#1}, #2 (#3)}
\def \ptp#1#2#3{Prog. Theor. Phys. {\bf#1}, #2 (#3)}
\def \ptwaw{Plenary talk, XXVIII International Conference on High Energy
Physics, Warsaw, July 25--31, 1996}
\def \rmp#1#2#3{Rev. Mod. Phys. {\bf#1}, #2 (#3)}
\def \rp#1{~~~~~\ldots\ldots{\rm rp~}{#1}~~~~~}
\def \si90{25th International Conference on High Energy Physics, Singapore,
Aug. 2-8, 1990}
\def \slc87{{\it Proceedings of the Salt Lake City Meeting} (Division of
Particles and Fields, American Physical Society, Salt Lake City, Utah, 1987),
ed. by C. DeTar and J. S. Ball (World Scientific, Singapore, 1987)}
\def \slac89{{\it Proceedings of the XIVth International Symposium on
Lepton and Photon Interactions,} Stanford, California, 1989, edited by M.
Riordan (World Scientific, Singapore, 1990)}
\def \smass82{{\it Proceedings of the 1982 DPF Summer Study on Elementary
Particle Physics and Future Facilities}, Snowmass, Colorado, edited by R.
Donaldson, R. Gustafson, and F. Paige (World Scientific, Singapore, 1982)}
\def \smass90{{\it Research Directions for the Decade} (Proceedings of the
1990 Summer Study on High Energy Physics, June 25--July 13, Snowmass, Colorado),
edited by E. L. Berger (World Scientific, Singapore, 1992)}
\def \stone{{\it $B$ Decays} (Revised 2nd Edition), edited by S. Stone
World Scientific, Singapore, 1994)}
\def \tasi90{{\it Testing the Standard Model} (Proceedings of the 1990
Theoretical Advanced Study Institute in Elementary Particle Physics, Boulder,
Colorado, 3--27 June, 1990), edited by M. Cveti\v{c} and P. Langacker
(World Scientific, Singapore, 1991)}
\def \waw{XXVIII International Conference on High Energy
Physics, Warsaw, July 25--31, 1996}
\def \yaf#1#2#3#4{Yad. Fiz. {\bf#1}, #2 (#3) [Sov. J. Nucl. Phys. {\bf #1},
#4 (#3)]}
\def \zhetf#1#2#3#4#5#6{Zh. Eksp. Teor. Fiz. {\bf #1}, #2 (#3) [Sov. Phys. -
JETP {\bf #4}, #5 (#6)]}
\def \zpc#1#2#3{Zeit. Phys. C {\bf#1}, #2 (#3)}
\def \zpd#1#2#3{Zeit. Phys. D {\bf#1}, #2 (#3)}

\end{document}